# Optical vortex dichroism in chiral particles

Kayn A. Forbes* and Garth A. Jones

*School of Chemistry, University of East Anglia, Norwich, Norfolk, NR4 7TJ, United Kingdom*

Circular dichroism is the differential rate of absorption of right- and left-handed circularly polarized light by chiral particles. Optical vortices which convey orbital angular momentum (OAM) possess a chirality associated with the clockwise or anti-clockwise twisting of their wavefront. Here it is highlighted that both oriented and randomly oriented chiral particles absorb photons from twisted beams at different rates depending on whether the vortex twists to the right or the left through a dipole coupling scheme. This is in contrast to previous studies that investigated dipole couplings with vortex modes in the paraxial approximation and showed no such chiral sensitivity to the vortex handedness: only in oriented media where electric quadrupole coupling contributes to optical activity effects due to absorption does such a mechanism exist for paraxial vortices. The distinct difference in the scheme highlighted in this work is that longitudinal fields are taken account of and these allow for OAM transfer even during dipole interactions. Due to the vortex dichroism persisting in randomly oriented collections of chiral particles, the mechanism has a distinct advantage in its potential applicability in chemical and biochemical applications where the systems under study are invariably in the liquid phase. Additionally, the result is put into context in terms of the quantifiable *optical chirality*, highlighting that optical OAM can in fact increase the optical chirality density of an electromagnetic field.

## I. INTRODUCTION

Natural optical activity is the differential interaction of a chiral particle - a chiral molecule for example - to the handedness of circularly-polarized light (CPL) [1]. Circular dichroism (CD) is a type of natural optical activity that specifically relates to a differential rate of absorption of CPL. Fundamentally all natural optical activity effects are due to the chiroptical interplay between the left- and right-handed nature of chiral particles $L(\xi)/R(\xi)$ and the left- and right-handed rotations of the electromagnetic field vectors in CPL $L(\sigma)/R(-\sigma)$ (see FIG. 1), this optical helicity is denoted by $\sigma = \pm 1$, respectively, and stems from the intrinsic spin angular momentum (SAM) $\sigma \hbar$ of photons.

Biomolecules are invariably chiral and therefore optically active, and the importance of chiroptical spectroscopies are at the forefront in determing these structures and their functionalities [2]. Indeed, CD methods have been well utilized in determing the secondary structure of proteins [3,4]. Raman optical activity (ROA) is a widely used technique in determining the chiral molecular structures and motions of viruses, carbohydrates, proteins, and structures even as large as insulin [5–7]. Beyond natural materials, chiroptical spectroscopies of fabricated nanostructures are an important facet of the fields of plasmonics and metamaterials [8–10].

*k.forbes@uea.ac.uk

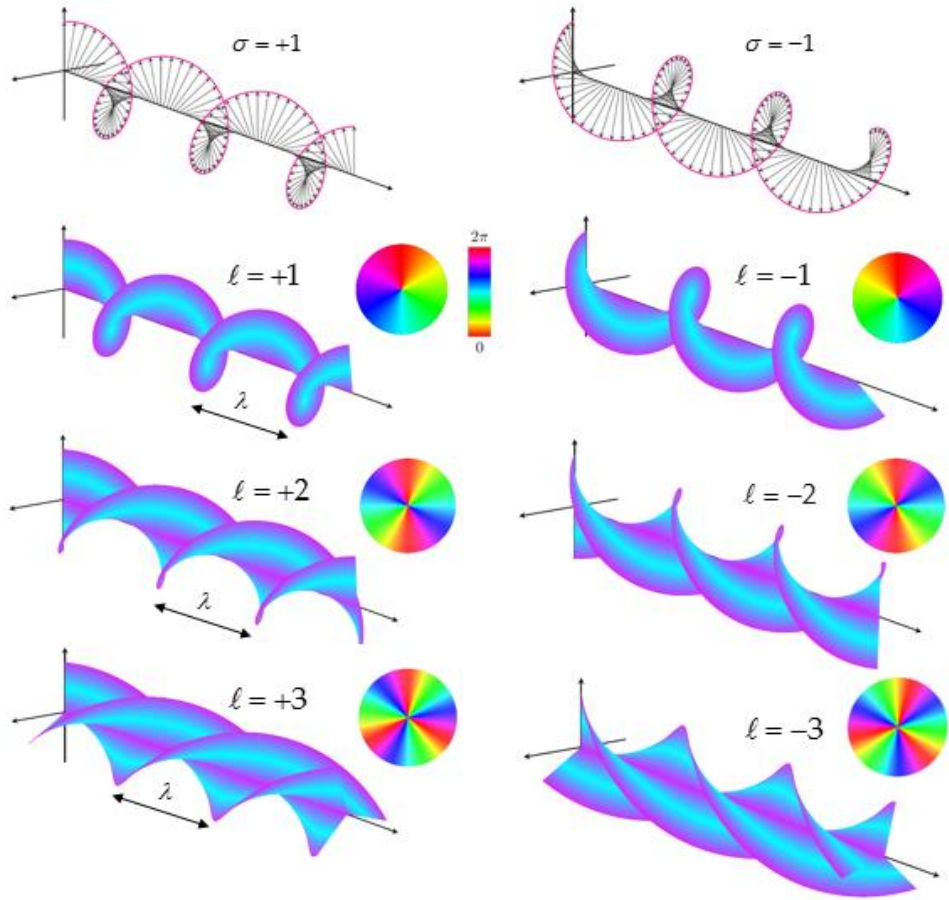

**FIG. 1** Chirality of light: the chirality of circular-polarization stems from the electromagnetic field vectors tracing out a helix as the light propagates; the chirality of an optical vortex stems from the helical structure that the wavefront traces out as the beam propagates. The optical helicity or chirality associated with lights polarization is an intrinsic property stemming from SAM; the vortex chirality is a spatial property of the beam stemming from its OAM (Figure reproduced with permission from [11]).

Optical vortices (or twisted light) are well-known and well-utilized in the physics community, particularly in optical manipulation; communication and information transfer; and imaging, to name a few [12]. The most common vortex mode implemented is the Laguerre-Gaussian (LG), a solution to the *paraxial* wave equation in cylindrical coordinates. The key property of optical vortices is their ability to convey orbital angular momentum (OAM) – this stems from the fact they propagate with a helical phase $e^{i\ell\phi}$, where $\ell \in \mathbb{Z}$ is known as the topological charge and $\phi$ is the azimuthal angle; optical vortices with $+|\ell|$ are left-handed, those with $-|\ell|$ are right-handed (See FIG.1). For beams propagating paraxially (that is, the wave vector makes a small inclination to the optical axis), the spin (polarization) and orbital (spatial) degrees of freedom are legitamately seperable, and individual photons may posess discrete $\ell\hbar$ units of OAM and $\sigma\hbar$ units of SAM [13,14]. The application of optical vortices in spectroscopic situations is a relatively new but rapidly growing research area, particularly in atomic optics [15] and chiral spectroscopies [11]. Interest in the latter is due to the fact optical vortices are chiral, in an analogous fashion to CPL (see FIG. 1), and so it is natural to ask whether (non-mechanical) light-matter interactions can be sensitive to the whether a vortex twists to the left or to the right.

At this point it is worth briefly mentioning an emerging application of the chirality of optical vortices to fabricate chiral micro- and nanostructures through purely mechanical means [16].

A comprehensive review of the field of chirality, optical activity and vortex light can be found in Ref [11]. The first study to address whether chiral molecules would respond differently to the handedness of an optical vortex considered paraxial LG vortex beams interacting in the dipole approximation (both electric and magnetic), concluding that neither oriented or randomly-oriented chiral molecules would show differential absorption of $+|\ell|$ and $-|\ell|$ photons [17]. More recent studies however highlighted how chiral molecules can in fact show differential interactions with paraxial vortices, but only through electric quadrupole (and higher multipole) interactions with the field [18,19], and in the specific case of absorption the individual material components have to be oriented with respect to the optical axis of the input beam. What all these studies have in common however is that they assume the input beam is well-described as a paraxial vortex mode and fully transverse with respect to the direction of propagation. Under specific situations, however, longitudinal (in the direction of beam propagation) fields can become highly important, such as when the fields are strongly-focused [20] or when specific angular momentum combinations are employed [21,22]. By accounting for longitudinal fields we highlight here a vortex dichroism (VD) which persists even for isotropic chiral particles in the dipole approximation. Compared to schemes where the chiral particles must exhibit a degree of orientational order, the VD mechanism outlined in this work that persists in orientationally averaged chiral particles has a much larger scope of potential applicability in chemical and biochemical chiral spectroscopies as the systems under study are invariably in solution and liquid phases.

## II. ABSORPTION OF TWISTED LIGHT BY CHIRAL PARTICLES

In the theory of quantum electrodynamics (QED) [23–25] the coupling of light and matter is represented by the interaction Hamiltonian $H_{int}$, which in the Power-Zienau-Woolley (PZW) formulation is given in a multipolar form for a particle $\xi$ positioned at $R_\xi$ [26]:

$$H_{int}(\xi) = -\varepsilon_0^{-1}\mu_i(\xi)d_i^\perp(R_\xi) - m_i(\xi)b_i(R_\xi) - \varepsilon_0^{-1}Q_{ij}(\xi)\nabla_i d_j^\perp(R_\xi) - ...,\qquad(1)$$

where $\boldsymbol{\mu}(\xi)$ is the electric dipole transition moment operator, $\boldsymbol{m}(\xi)$ the magnetic dipole, and $Q_{ij}(\xi)$ electric quadrupole; $d^\perp(R_\xi)$ and $b(R_\xi)$ are the electric displacement field and magnetic field operators, respectively (both transverse to the Poynting vector); we use standard suffix notation for tensor quantities and imply the Einstein summation convention for repeated indices throughout (i.e. $a_i b_i = \boldsymbol{a}\cdot\boldsymbol{b}$).

The electric displacement field mode expansion for circularly-polarized Laguerre-Gaussian (LG) beams in the long Rayleigh range is given by [22]:

$$d_i^\perp(\mathbf{r}) = \sum_{k,\sigma,\ell,p} \frac{\Omega}{\sqrt{2}} \left[ \left\{ (\hat{x} + i\sigma\hat{y})_i + \frac{i}{k} \left( \frac{\partial}{\partial r} - \ell\sigma \frac{1}{r} \right) e^{i\sigma\phi} \hat{z}_i \right\} f_{|\ell|,p}(r) a_{|\ell|,p}^{(\sigma)}(k\hat{z}) e^{i(kz+\ell\phi)} - H.c. \right], \qquad (2)$$

where $\Omega = i\left( \hbar c k \varepsilon_0 / 2 A_{\ell,p}^2 V \right)^{1/2}$ is the normalization constant for LG modes, with $V$ the quantization volume; $f_{|\ell|,p}(r)$ is a radial distribution function (see S.I.); $a_{|\ell|,p}^{(\sigma)}(k\hat{z})$ is the annihilation operator; the exponential terms $e^{ikz}$, $e^{i\ell\phi}$, and $e^{i\sigma\phi}$ are phase factors; and *H.c.* stands for Hermitian conjugate. The magnetic field CPL LG mode operator in the long Rayleigh range is

$$b_i(\mathbf{r}) = \sum_{k,\sigma,\ell,p} \frac{\Omega}{\sqrt{2}\varepsilon_0 c} \left[ \left\{ (\hat{y} - i\sigma\hat{x})_i + \frac{1}{k} \left( \sigma\frac{\partial}{\partial r} - \frac{\ell}{r} \right) e^{i\sigma\phi} \hat{z}_i \right\} f_{|\ell|,p}(r) a_{|\ell|,p}^{(\sigma)}(k\hat{z}) e^{i(kz+\ell\phi)} - H.c. \right]. \qquad (3)$$

In Eq. (2) and Eq. (3), the part of the total fields that depend on the transverse coordinates $\hat{x}$ and $\hat{y}$ are known in terminology originating from Lax *et al.* [27] as the zeroth-order transverse fields, the terms dependent on $\hat{z}$ are the first-order longitudinal fields, generally neglected for paraxial vortex modes such as LG. The relative magnitude of the longitudinal field compared to the transverse fields is weighted by the paraxial paramter $(kw_0)^{-1}$, where $w_0$ is the beam waist at $z = 0$. For general non-OAM posessing laser modes (e.g. a Gaussian mode) longitudinal fields only become important for very highly-focused non-paraxial laser fields and can be safely neglected for paraxial modes [20]. However, this is not the case for OAM-possessing paraxial optical vortex light, and in general first-order longitudinal fields of optical vortices should be included for beams with larger values of $kw_0$ than would be necessary for non-OAM paraxial posessing modes [21,22].

Most optical interactions with matter can be solely described by E1 couplings (the first term on the right-hand side of Eq. (1)): the electic dipole approximation. However, the origin of natural optical activity stems from the *interferences* between electric dipole couplings (E1) with both magnetic dipole (M1) and electric quadrupole (E2) couplings: E1M1 and E1E2 [28]. For CD in particular, isotropic collections of chiral particles (e.g. those randomly oriented such as in a liquid) only produce optical activity through the E1M1 mechanism: E1E2 contributes to oriented systems of molecules but averages to zero for tumbling systems [29].

As mentioned in Sec I., the first study in the field of optical activity and twisted light predicted via theoretical calculations that chiral molecules (both oriented and randomly-oriented) subjected to an optical vortex would show no differential rate of absorption with respect to the vortex handedness [17]. However, that study only included the purely zeroth-order transverse fields of the Laguerre-Gaussian modes Eq. (2) and Eq. (3). In this study we highlight how inclusion of the longitudinal fields does in fact allow for a vortex dichroism of twisted photons for both ordered and isotropic chiral molecules through the E1M1 mechanism.

For CD the initial and final states of the total light-matter system are given by the following kets, respectively: $|I\rangle = |E_0\rangle |n(k,\sigma,\ell,p)\rangle$ and $|F\rangle = |E_\alpha\rangle |n-1(k,\sigma,\ell,p)\rangle$. That is, a chiral particle in

the initial state $|E_0\rangle$ absorbs a photon from a single mode $(k, \sigma, \ell, p)$ input laser with occupation number $n$, reducing the occupaton of the mode to $n-1$ and resulting in the particle being in the excited state $|E_\alpha\rangle$. CD involves the absorption of a single photon, and so first-order time-dependent pertubation theory $M_{FI} = \langle F | H_{int}(\xi) | I \rangle$ yields the matrix element (or quantum amplitude) for the process:

$$M_{FI} = -\frac{\Omega\sqrt{n}}{\sqrt{2}\varepsilon_0}\left[\left\{(\hat{x}+i\sigma\hat{y})_i + \frac{i}{k}\left(\frac{\partial}{\partial r} - \frac{\ell\sigma}{r}\right)e^{i\sigma\phi}\hat{z}_i\right\}\mu_i^{\alpha 0}\right.$$
$$\left. + \frac{1}{c}\left\{(\hat{y}-i\sigma\hat{x})_i + \frac{1}{k}\left(\sigma\frac{\partial}{\partial r} - \frac{\ell}{r}\right)e^{i\sigma\phi}\hat{z}_i\right\}m_i^{\alpha 0}\right]f\,e^{i(kz+\ell\phi)}, \tag{4}$$

where for notational brevity we drop the dependencies of the radial function $f$. To calculate the rate $\Gamma$ of photon absorption we require Fermi's rate rule: $\Gamma = 2\pi\hbar^{-1}|M_{FI}|^2\rho$ where $\rho$ is the density of final states. When taking the modulus square of Eq. (4) as required by the Fermi rule it is evident that three distinct terms will be produced: $\mu\mu$, $mm$ and $\mu m$, the latter of these are the interference terms between electric and magnetic transition dipole moments (E1M1) and these are what are responsible for the differential effects observed in optical activity [1,23]. The pure electric dipole E1E1 and pure magnetic dipole terms M1M1 are equivalent for either enantiomer – this is clearly obvious from parity considerations alone [28] – and therefore are neglected from now on apart form when discussing Kuhn's dissymmetry factor.

In the fields of chemistry, biochemistry, and molecular spectroscopy the particles under study are often in the condensed liquid phase and exhibit no correlations between one another. Chiral molecules in fluids are generally randomly oriented with respect to the laboratory frame of reference, and to account for this isotropic system we must carry out an orientational average of the molecules (see S.I. for the result that pertains to oriented chiral systems). The second-rank molecular tensor average is easily carried out using standard techniques [30], namely $\langle a_i b_j \rangle = 3^{-1}\delta_{ij}\boldsymbol{a}\cdot\boldsymbol{b}$, where angular brackets denote a rotationally averaged quantity. Taking all of the above into account, the rate is given by

$$\langle\Gamma\rangle \propto \langle|M_{FI}|^2\rangle \propto -N_r\frac{I}{3\varepsilon_0 c^2}\left[\sigma f^2 + \frac{1}{2k^2}\left(\sigma f'^2 + \frac{\ell^2\sigma}{r^2}f^2 - \frac{2\ell}{r}ff'\right)\right]\boldsymbol{\mu}^{0\alpha}\cdot i\boldsymbol{m}^{\alpha 0}, \tag{5}$$

where $N_r$ is the number of chiral particles at a position $r$, the beam irradiance is given by $I = n\hbar c^2 k \big/ A_{\tilde{c},p}^2 V$ and $\mu_i^{0\alpha} = \bar{\mu}_i^{\alpha 0}$ is pure a real polar vector, but $m_i^{0\alpha} = -\bar{m}_i^{0\alpha}$ and $m_i^{0\alpha} = -m_i^{\alpha 0}$ is a purely imaginary axial vector.

## III. CIRCULAR DICHROISM WITH OPTICAL VORTICES

The circular rate differential for CD, i.e. the difference of absorption between left $(\sigma = +1)$ and right $(\sigma = -1)$ handed circular polarization is easily derived from Eq. (5):

$$\left\langle \Gamma^{(L)} \right\rangle - \left\langle \Gamma^{(R)} \right\rangle \propto N_r \frac{2I}{3\varepsilon_0 c^2}\left[ f^2 + \frac{1}{2k^2}\left( f'^2 + \frac{\ell^2}{r^2} f^2 \right) \right] R^{\alpha 0}, \tag{6}$$

where we now make use of the optical rotatory tensor defined as usual by [31,32]

$$R_{ij}^{\alpha 0} = \mathrm{Im}\left\langle E_0 \left| \mu_i \right| E_\alpha \right\rangle\left\langle E_\alpha \left| m_j \right| E_0 \right\rangle = \mathrm{Im}\, \mu_i^{0\alpha} m_j^{\alpha 0}, \tag{7}$$

which in its rotationally averaged form is the pseudoscalar optical rotatory strength $R^{\alpha 0}$. Importantly Eq. (6) can easily be shown to give the well-known Kuhn's dissymmetry factor for randomly-oriented chiral molecules:

$$g = \frac{\left\langle \Gamma^{(L)} \right\rangle - \left\langle \Gamma^{(R)} \right\rangle}{\frac{1}{2}\left( \left\langle \Gamma^{(L)} \right\rangle + \left\langle \Gamma^{(R)} \right\rangle \right)} \propto \frac{4}{c}\frac{\left[ f^2 + \frac{1}{2k^2}\left( f'^2 + \frac{\ell^2}{r^2} f^2 \right) \right] R^{\alpha 0}}{\frac{1}{2}\left[ 2f^2 + \frac{1}{k^2}\left( f'^2 + \frac{\ell^2}{r^2} f^2 \right) \right]\left| \boldsymbol{\mu}^{\alpha 0} \right|^2} = \frac{4}{c}\frac{R^{\alpha 0}}{\left| \boldsymbol{\mu}^{\alpha 0} \right|^2}. \tag{8}$$

The dissymmetry factor Eq. (8) is clearly position-independent even for a structured beam. The rate differential Eq. (6) for $|\ell| = 1,2$ is plotted in FIG. 2.

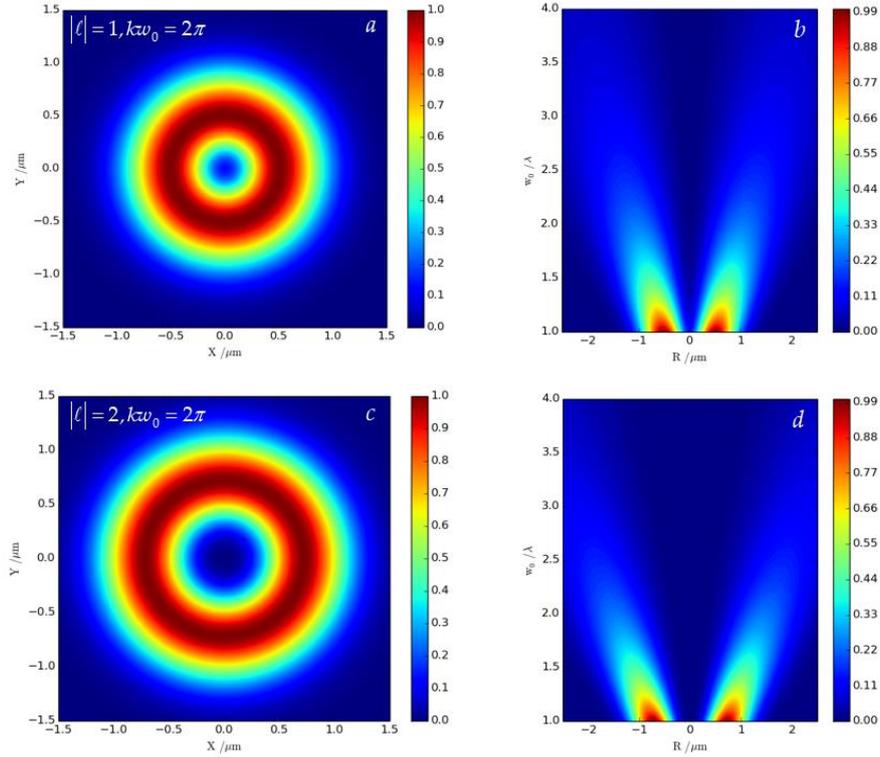

**FIG. 2** Circular dichroism (CD) Eq. (6). a) $|\ell|=1$, $w_0=\lambda$ b) $|\ell|=1$ for values of $2\pi \geq kw_0 \geq 8\pi$ c) $|\ell|=2$, $w_0=\lambda$ d) $|\ell|=2$ for values of $2\pi \geq kw_0 \geq 8\pi$.

Firstly we must make clear that this CD does not depend on the sign of $\ell$, i.e. it is not sensitive to the handedness of the input vortex. The peak signal intensity at any given location is larger for smaller values of $kw_0$ for a given input beam power, this simply indicates the energy of the beam being spread into a smaller area. However, we also see that for a small range of $kw_0$ the circular vortex differential in the core of an $|\ell|=1$ mode is not actually zero as would be expected for a doughnut-shaped vortex mode. This stems from the decreasing value of $kw_0$ leading to larger longitudinal field components, contributions of which yield an on-axis intensity for $|\ell|=1$. For values of $kw_0 < 2\pi$ the on-axis intensity of the $|\ell|=1$ continues to increase and can even become larger than the transverse components of the field; for such values of $kw_0$ there can even exist an on-axis intensity for $|\ell|=2$. However, such extreme focusing scenarios are beyond the validity of the paraxial wave equation – see Sec. V. for further discussion.

The magnitude of the CD differential is highly-dependent on the location of the chiral particle in the beam; note that all values of the CD differential are of the same sign for any given $r$ and so all signals effectively positively contribute to the total differential rate. Note that it is not the wavelength that necessarily dictates the distributions in FIG 2. but rather the value of $kw_0$.

## IV. VORTEX DICHROISM

Bearing in mind the two different forms of optical handedness (FIG. 1), the CPL handedness exhibited via $\sigma = \pm 1$ and the optical vortex handedness exhibited via $\pm |\ell|$, clearly there are a number of distinct scenarios of where this chirality can engage in chiroptical effects. For a thorough discussion we again refer the reader to Ref. [11], however for our purposes here we note that in theory chiroptical effects may depend on the CPL handedness as in CD, the vortex handedness through the sign of $\ell$, or both forms of handedness through the product $\ell \cdot \sigma$. The latter of these lead to so-called circular-vortex differential effects, whilst those solely dependent on the vortex handedness would be vortex differential effects, and in specific case of absorption it may be termed vortex dichroism (VD).

Inspecting Eq. (5) it can be seen that the last term in rounded brackets which is linearly dependent on $\ell$ can also produce a differential rate, but in contrast to CD this one manifests for $+|\ell| - (-|\ell|)$. This VD differential is obtained as

$$\left\langle \Gamma_{|\ell|}^{|\sigma|} \right\rangle - \left\langle \Gamma_{-|\ell|}^{|\sigma|} \right\rangle \propto \frac{2 N_r |\ell|}{3 \varepsilon_0 c^2 k^2 r} f\!f \mathcal{R}^{\alpha 0}.$$
(9)

Importantly the VD differential Eq. (9) is independent of the handedness of the input circular polarization. Throughout the derivation so far we have assumed the input beam to be circularly polarized for the sake of generality – we now therefore set $\sigma = 0$ and see that we still produce a VD differential (see S.I.), namely

$$\left\langle \Gamma_{|\ell|}^{\sigma = 0} \right\rangle - \left\langle \Gamma_{-|\ell|}^{\sigma = 0} \right\rangle \propto \frac{N_r I |\ell|}{3 \varepsilon_0 c^2 k^2 r} f\!f \mathcal{R}^{\alpha 0}.$$
(10)

In other words, $\left\langle \Gamma_{|\ell|}^{|\sigma|} \right\rangle - \left\langle \Gamma_{-|\ell|}^{|\sigma|} \right\rangle = 2 \left( \left\langle \Gamma_{|\ell|}^{\sigma = 0} \right\rangle - \left\langle \Gamma_{-|\ell|}^{\sigma = 0} \right\rangle \right)$. The VD differential rate Eq. (9) is plotted in FIG. 3 for $|\ell| = 1, 2$. Although the strength of the CD signal varies with the radial position of the chiral particle in the beam in FIG. 2, the magnitude is always the same sign, positive in this case (negative if taking $\left\langle \Gamma^{(R)} \right\rangle - \left\langle \Gamma^{(L)} \right\rangle$). However, as FIG. 3 highlights, the VD signal is likewise highly-position dependent, but can take on negative as well as positive values.

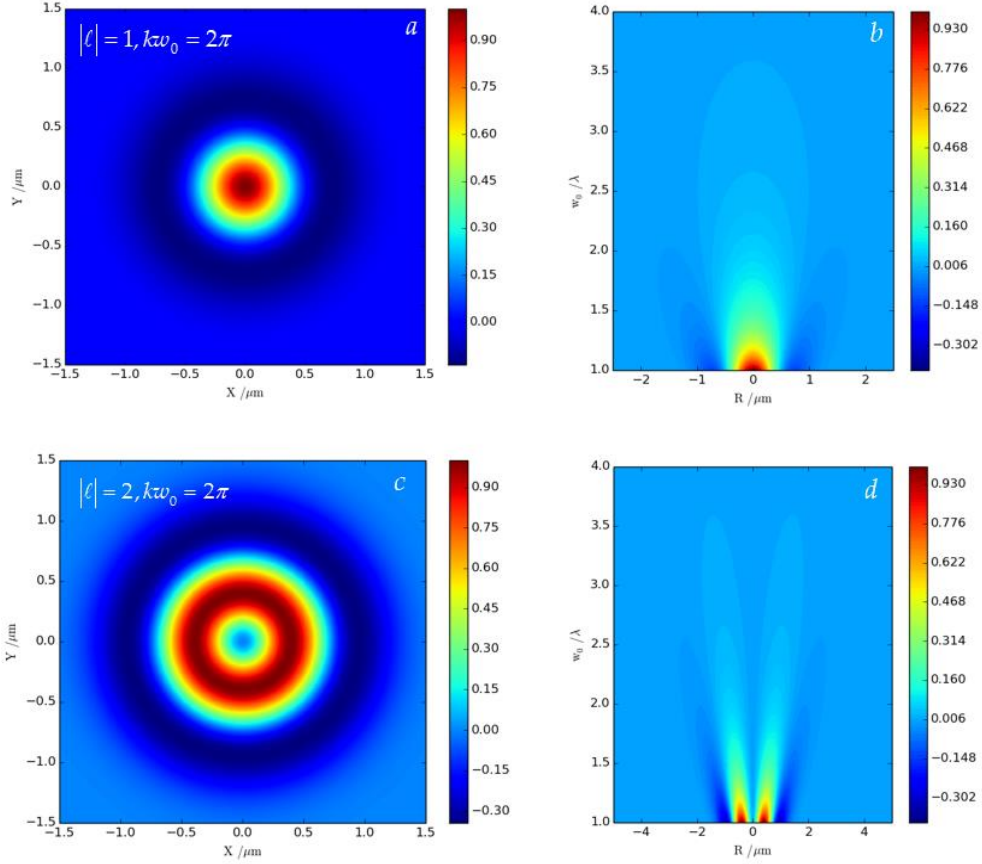

**FIG. 3** Vortex dichroism (VD) Eq. (9). $\left(|\sigma|=1\right)$ a) $|\ell|=1$, $w_0 = \lambda$ b) $|\ell|=1$ for values of $2\pi \geq kw_0 \geq 8\pi$ c) $|\ell|=2$, $w_0 = \lambda$ d) $|\ell|=2$ for values of $2\pi \geq kw_0 \geq 8\pi$

The integrated signals of FIG. 3 are highly dependent on the size and number ($N_r$) of the individual chiral particles under study (See Sec. V for further discussion). The VD effect does not give the standard Kuhn's dissymmetry factor for isotropic chiral particles either and is plotted in FIG. 4 for $|\ell|=1,2$:

$$ g = \frac{4|\ell| \, ff \, \mathcal{R}^{a0}}{crk^2 \left[ f^2 + \frac{1}{2k^2}\left( f'^2 + \frac{\ell^2}{r^2} f^2 \right) \right] |\boldsymbol{\mu}^{a0}|^2}. \tag{11} $$

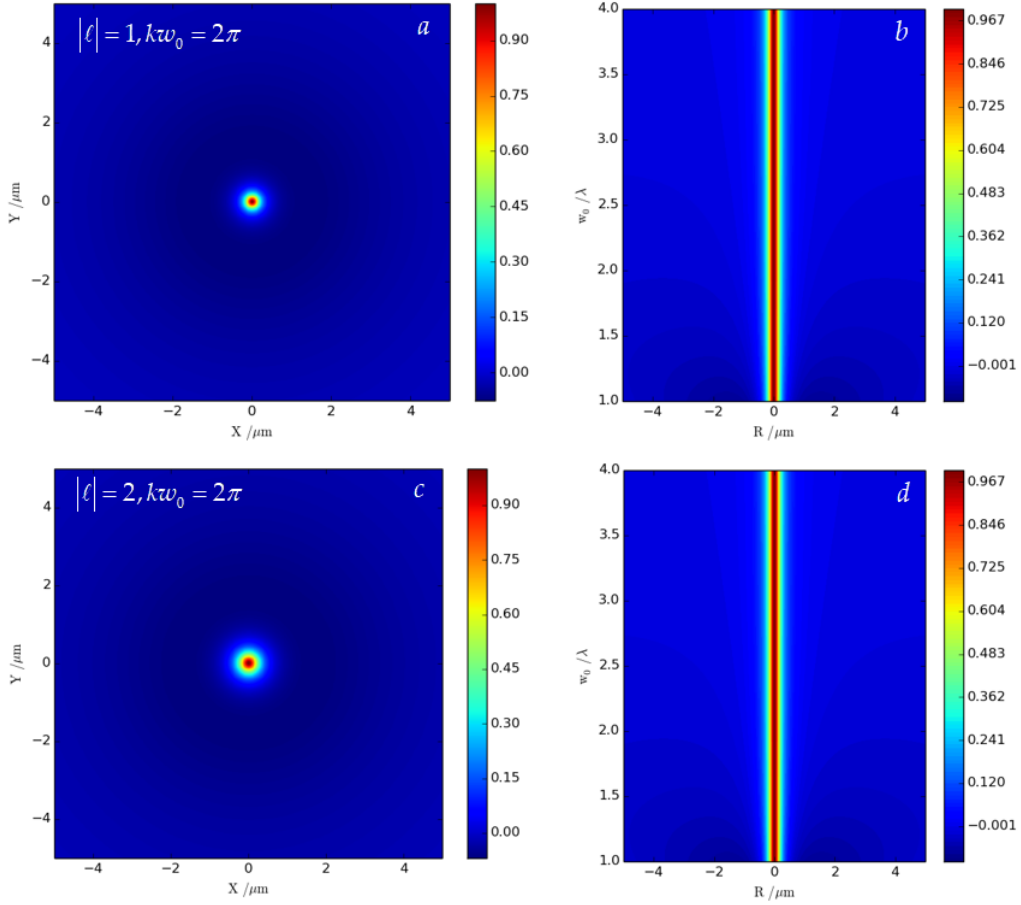

**FIG. 4** Kuhn's dissymmetry factor Eq. (11). a) $|\ell| = 1$, $w_0 = \lambda$  b) $|\ell| = 1$ for values of $2\pi \geq k w_0 \geq 8\pi$  c) $|\ell| = 2$, $w_0 = \lambda$  d) $|\ell| = 2$ for values of $2\pi \geq k w_0 \geq 8\pi$. Compared to previous figures, the ones here are plotted with a wider range of *X* and *Y* values to show the negative ring of signal intensity.

## V. ANALYSIS AND DISCUSSION

The initial theoretical study we have mentioned previously that looked at E1M1 interactions with twisted light used only the zeroth-order transverse parts of Eq. (2) and Eq. (3), concluding that chiral molecules do not interact with the sign of $\ell$ in a chiroptical fashion [17]. This study was followed up by experimental work that seemingly vindicated this theoretical prediction [33,34]. Whilst Araoka *et al.* studied their chiral material system under the influence of a weakly-focused LG vortex, Löffler *et al.* used both weakly- and strongly focused LG modes. The experimental result that the sign of $\ell$ plays no role in CD under weak-focusing can be explained by the initial theory of Andrews *et al.* [17] or that in Sec. III and Sec. IV of this work, namely that under such conditions the CD differential has no dependence on $\ell$ and that the longitudinal fields responsible for VD are insignificant as $w_0 \gg \lambda$.

We believe a potential reason the Löffler *et al* strongly-focused beam experiment failed to observe differential effects dependent on $\ell$ is due to the dissymmetry factor they used to measure an influence of OAM on CD. Their definition is

$$g_{\text{OAM}} = \frac{\left(\Gamma_{+|\ell|}^{+|\sigma|} - \Gamma_{+|\ell|}^{-|\sigma|}\right) - \left(\Gamma_{-|\ell|}^{+|\sigma|} - \Gamma_{-|\ell|}^{-|\sigma|}\right)}{\left(\Gamma_{+|\ell|}^{+|\sigma|} - \Gamma_{+|\ell|}^{-|\sigma|}\right) + \left(\Gamma_{-|\ell|}^{+|\sigma|} - \Gamma_{-|\ell|}^{-|\sigma|}\right)}. \tag{12}$$

Inserting our Eq. (5) into this definition automatically leads to a null result for any dependence on $\ell$ and also for each individual differential measurement in Eq. (12). The necessary observable should use a fixed value for $\sigma$ or alternatively $\sigma = 0$, as highlighted in Sec. IV, one such example being:

$$g_{\text{OAM}} = \frac{\Gamma_{+|\ell|}^{|\sigma|} - \Gamma_{-|\ell|}^{|\sigma|}}{\Gamma_{+|\ell|}^{|\sigma|} + \Gamma_{-|\ell|}^{|\sigma|}}, \tag{13}$$

which is essentially the dissymmetry factor plotted in FIG. 4.

Looking at FIG. 3b and 3d it is clear the VD differential increases as the value of $kw_0$ gets smaller. This is to be expected as the VD effect itself stems from longitudinal fields, which are correlated to the degree of focusing, though for vortices are also sensitive to input angular momenta configurations. For example, we have seen in Sec. IV that the VD differential is twice as large if the input beam is circularly polarized compared to linearly polarized.

Whilst higher-order transverse and longitudinal components to the mode expansions Eq. (2) and Eq. (3) can be derived using the Maxwell-Ampere and Faraday laws, LG modes are fundamentally solutions to the paraxial wave equation, and so are always bound to the well-known approximations associated with it. As such, any contribution to Eq. (2) and Eq. (3) from these higher-order fields only become relevant in situations where $w_0 < \lambda$ and a paraxial wave solution (such as the LG mode) is not justifiable in this regime. Specifically, we can expect our theory to work well both qualitatively and quantitatively so long as $w_0 \geq \lambda$ [35,36], however using our theory to go below this lower bound will introduce quantitative errors. We can however predict with confidence that the VD signal will only get larger for sub-wavelength focusing where $w_0 \leq \lambda$, however a more refined theory which specifically accounts for non-paraxial fields should be used to yield accurate results with respect to the quantitative magnitudes involved. Such methods could be those that explicitly account for high-NA focusing [20,37], for example, or alternatively non-paraxial solutions to the full Helmholtz equation should be used, such as Bessel modes (or the non-paraxial form of LG modes) [38]. Further interesting properties may arise if the field is highly non-paraxial, as it is known that an on axis intensity can exist in even the $|\ell| = 2$ case for such strongly-focussed scenarios when second-order transverse components of the field can produce observable effects [39].

Another important issue we draw attention to is the unique scale-dependent nature of VD. Because the VD signal can take on both positive and negative signals, the size and number of any chiral particles within the interaction volume is important. One can imagine scenarios with specifically sized particles where the signal is essentially zero; alternatively, increasing the value of $|\ell|$ could lead to enhanced signals as a larger number of particles of a given size could fit in the high signal intensity region compared to a lower value of $|\ell|$. Because the CD signal is always of the same sign, such a scheme is not possible. This difference in behaviour between CD and VD is another classic motif of chiroptical effects with optical vortices [11]. CD stems from the intrinsic property of circular polarization; VD stems from the OAM of a vortex which is a spatial property of the beam, and thus any measurement of the chirality associated to it is likewise scale dependent. This is no different to the geometrical chirality of molecules; a small chiral molecule is no less chiral than a large one; both their chiral nature is exhibited to varying degrees depending on what they are specifically interacting with. It should therefore be of no surprise that the ability to exhibit a chiroptical effect in small chiral particles using an optical vortex like we have shown here with VD requires small values of $w_0$ on the order $\lambda$. A similar logic of matching the size of an optical vortex to the material dimensions has recently highlighted a differential scattering effect for chiral microstructures which exhibit an acute sensitivity to the magnitude and sign of $\ell$ [40].

Finally, an important experimental technique of carrying out optical activity studies is the ability to modulate between left and right circular polarizations. With respect to experimentally observing the VD effect outlined here it is pertinent to note the timely technical breakthrough by J.-F. Bisson *et al.* of the ability to modulate between optical vortices with different signs of $\ell$ [41].

## VI. OPTICAL CHIRALITY

A different way of interpreting the results of Sec. III and Sec. IV is through the so-called optical chirality density $\chi$. Originally introduced by Lipkin [42] and brought to prominence by Tang and Cohen [43], crudely put it quantifies how chiral an electromagnetic field is. The optical chirality density in free space for a monochromatic beam may be given by [44]

$$\chi = \omega^2 \left( -\int d^{\perp} dt \right) \cdot \boldsymbol{b}. \tag{14}$$

In their paper, Coles and Andrews [44] state that the optical chirality Eq. (14) is independent of any factors pertaining to the optical orbital angular momentum. Their analysis, fully correct when only zeroth-order transverse fields are accounted for, however, neglects the first-order longitudinal fields which we have seen lead to the VD chiroptical effect. Inserting Eq. (2) and Eq. (3), which include the longitudinal fields, into Eq. (14) we discover that the optical chirality density *does* in fact depend on optical OAM. Specifically

$$\chi = \sum_{k,\sigma,\ell,p} \left( \frac{n\hbar c k^2}{A_{\ell,p}^2 V} \right) \left[ \sigma ff + \frac{1}{4k^2} \left( \sigma ff' + \frac{\ell^2 \sigma}{r^2} ff - \frac{2\ell}{r} ff' \right) \right]. \tag{15}$$

This equation tell us is that there is a non-zero optical chirality density at $z = 0$ which is made up from contributions dependent on both $\ell$ and $\sigma$, these individual contributions are plotted in FIG. 5. In Eq. (15) the first term in square brackets on the right-hand side is the standard zeroth-order paraxial contribution to the optical chirality density; the remaining terms (in round brackets) all originate from the longitudinal contributions to the fields. It is clear that Eq. (15) and FIG. 5 mirror the physics of those in the previous sections which were calculated using standard perturbative QED methods. A more obvious consequence of Eq. (15) however is that optical orbital angular momentum can lead to an enhanced optical chirality density in comparison to non-OAM possessing ($\ell = 0$) light, and this enhancement is proportional to the smallness of $kw_0$.

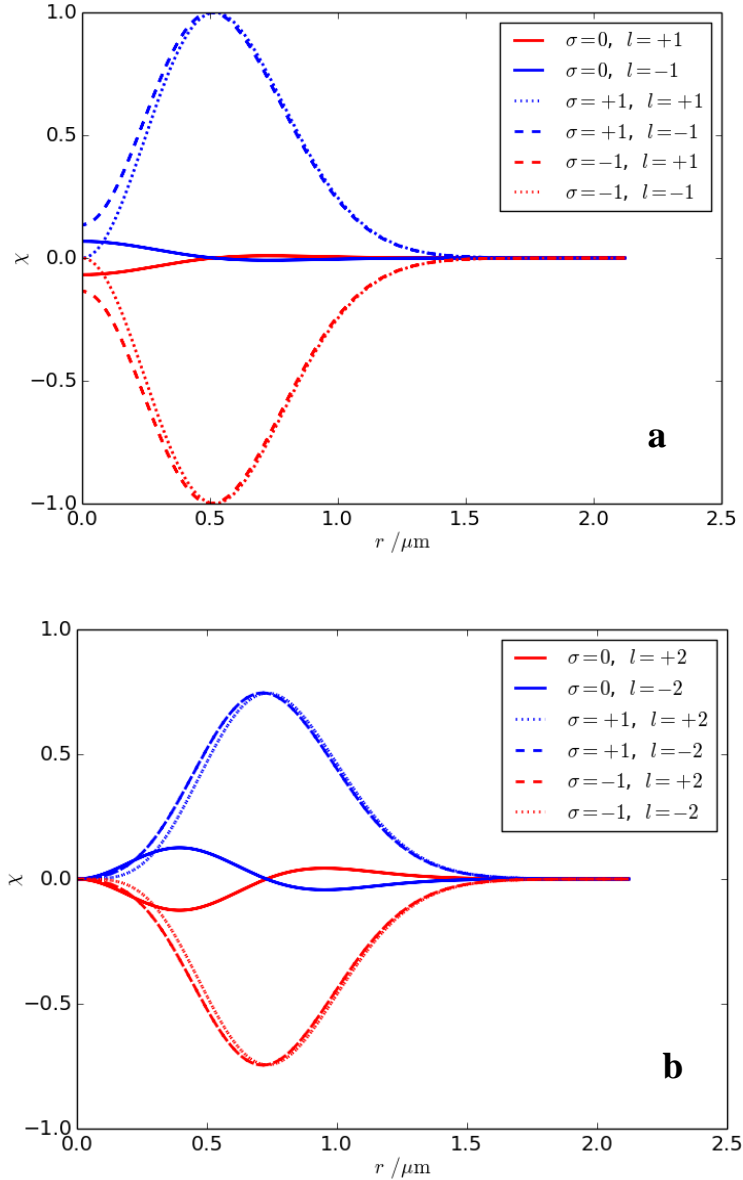

**FIG. 5** *Optical chirality* density Eq. (15) for $kw_0 = 2\pi$ and different combinations of OAM and SAM a) $|\sigma| = 0,1 \quad |\ell| = 1$ b) $|\sigma| = 0,1, |\ell| = 2$. Note the existence of an optical chirality density for optical vortices even when $\sigma = 0$.

The on-axis optical chirality density in Eq. (15) (and FIG. 5a) present for $\sigma = 0, \ell = \pm 1$ stems purely from longitudinal fields and was first noted by Rosales-Guzmán *et al.* for linearly-polarized Bessel beams [45], subsequently Woźniak *et al.* experimentally observed this on-axis optical chirality using focused $kw_0 \approx 8.7$ linearly-polarized LG modes interacting with a chiral plasmonic helix. It is worthwhile to note that our analysis here indicates that if the input beam was circularly-polarized $|\sigma| = 1$ then the effects in these two previous studies would be twice the size in magnitude. Furthermore, our results also extend these studies to values of $|\ell| > 1$.

## VII. CONCLUSION

Here we have highlighted a vortex dichroism exhibited by chiral particles that are both oriented and randomly oriented which stems from longitudinal fields of optical vortices through an E1M1 mechanism. This contrasts to previous work restricted to the purely zeroth-order transverse fields of a paraxial vortex in which no such mechanism is viable through the E1M1 route. Previously discovered mechanisms in the paraxial approximation required E2 couplings which vanish for randomly oriented chiral particles. It is important to note the underlying origin of both mechanisms is that the material is interacting with the transverse phase gradient of the input mode [46], i.e. the property responsible for the OAM.

Whilst previous chiroptical absorption mechanisms have been discovered for vortex light in oriented media, the importance of the result here is that it persists in fluids consisting of chiral particles, which has an acute importance in the field of optical activity and spectroscopies utilized in chemical and biochemical systems which are invariably in the liquid phase. There have been both a number of circular-vortex differential [19,47,48] and vortex differential [40,49–52] effects reported, though no vortex dichroism (absorption) effects in isotropic chiral molecular matter of the nature here has been reported thus far to the best of our knowledge. The underlying principles of this work can easily be extended to other types of optical activity, such as optical rotation or Rayleigh and Raman optical activity (scattering effects). To experimentally observe VD in small chiral particles the input optical vortex must be moderately to strongly-focused $kw_0 \leq 8\pi$, though may still be viable for larger values, and the differential under study should be of the form Eq. (13). Whilst the accuracy of our analysis is quantitatively limited to values of $w_0 > \lambda$, in fields which are focused further, i.e. $w_0 < \lambda$, the VD differential should only increase in magnitude, and for when $|\ell| = 2$ the effect will even occur on-axis due to the on-axis intensity that is known in this scenario.

Furthermore, we highlighted the agreement between our result derived using standard perturbative QED methods and that using the quantity known as *optical chirality*. This also allowed us to highlight that optical OAM can in fact increase the optical chirality density, contrary to earlier studies restricted to the paraxial regime.

## ACKNOWLEDGEMENT

KAF is grateful to the Leverhulme Trust for funding him through a Leverhulme Trust Early Career Fellowship (Grant Number ECF-2019-398).

---